\documentclass[conference]{IEEEtran}
\IEEEoverridecommandlockouts
\usepackage{cite}
\usepackage{float}
\usepackage{amsmath}
\usepackage{amssymb}
\usepackage{tabularx}
\usepackage{amsfonts}
\usepackage{algorithmic}
\usepackage{graphicx}
\usepackage{subcaption}
\usepackage{textcomp}
\usepackage{float}
\usepackage{wrapfig}
\usepackage{url}
\usepackage{threeparttable}
\usepackage[inkscapeformat=png]{svg}
\usepackage{xcolor}
\def\BibTeX{{\rm B\kern-.05em{\sc i\kern-.025em b}\kern-.08em
    T\kern-.1667em\lower.7ex\hbox{E}\kern-.125emX}}
\begin{document}
\title{System-level Testing of the Congestion Management Capability of a Hardware-Independent Optimal Power Flow Algorithm}

\author{\IEEEauthorblockN{Thomas Schwierz, Rajkumar Palaniappan, Oleksii Molodchyk, Christian Rehtanz}
\IEEEauthorblockA{
\textit{Institute of energy systems, energy efficiency and energy economics} \\
\textit{TU Dortmund University}\\
Dortmund, Germany\\
\{Thomas.Schwierz; Rajkumar.Palaniappan; Oleksii.Molodchyk; Christian.Rehtanz\}@tu-dortmund.de}
}

\maketitle

\begin{abstract}

The integration of distributed energy resources (DERs) into the electrical grid causes various challenges in the distribution grids. The complexity of smart grids as multi-domain energy systems requires innovative architectures and algorithms for system control. While these solutions are good on paper, several testing methods are required to test the applicability of components, functions and entire systems to the existing energy grids. In this paper, a full-scale low-voltage test setup in the Smart Grid Technology Lab (SGTL) at TU Dortmund University is used to evaluate the capability of an Optimal Power Flow Algorithm (OPF) to support voltage control, congestion management, and to provide redispatch to the higher grid levels. While conventional redispatch is commonly done preemptively, this paper analyses the possibility of providing redispatch to the higher voltage levels without taking the future grid state into consideration. The importance of this implementation is that the smart grid application used to execute the OPF is configured based on IEC 61850 data models, making the software independent of the hardware. Such standardised control algorithms are interoperable and can be implemented on any hardware that suits the requirements.

\end{abstract}


\section{Introduction}
Power systems have proven to be an integral part of modern society, and the electrical energy industry of today is changing due to several reasons. The planning and operation of electrical energy grids has become enormously complicated in recent years. With the integration of various new loads and distributed energy resources (DERs) into the grid, the power system industry is looking for new and innovative solutions to counter the challenges of increased grid loading, reverse power flows, voltage problems and low power reserves on the transmission system level, only to mention some of them \cite{Breker}. Simultaneously, new technologies and developments in Information and Communication Technology (ICT) infrastructure promise to enable the distribution system operators (DSO) with additional options to control the assets in their grid and address the dynamic changes in their grid. With these new technologies, innovative monitoring and control algorithms are proposed for active distribution grids \cite{Hu}. Whenever new algorithms are introduced in the literature, it is common practice to test them before implementing them in the field. The reason is simple: Many programming and algorithmic errors can be corrected, which may have been overlooked during the development phase.

Hardware-in-the-loop (HIL) testing methods are a common testing method in the present-day world since they include the domain of communication and enable the testing of hardware devices, such as inverters, or controllers in real-time. HIL testing usually entails a digital real-time simulated model and the hardware under test connected using analogue/digital inputs/outputs. In the context of the control of power systems, the power system is usually modelled on a real-time simulator (RTS) and communicates measurements and status values of the power system to the controller, whilst the controller sends outputs to the flexible assets of the simulated grids  \cite{Montoya}. This concept is called Controller-Hardware-in-the-loop (CHIL). In Power-Hardware-in-the-loop (PHIL) testing, the Input-Output-Interface of the RTS is connected to a power interface to emulate the currents and voltages as they occur in the power system. Whilst CHIL testing can be beneficial for individual controller testing, the transition of distribution grids from passive distribution grids to active distribution grids, also known as smart grids, demands a more integrated testing approach. Therefore, the approach of system-level validation has been developed \cite{Strasser}, \cite{1}. System-level validation approaches include multiple \textendash{} potentially HIL controlled \textendash{} components, communication, a model of or a physical realisation of a power grid, a visualisation and a control system. At TU Dortmund University, the Smart Grid Technologies Lab (SGTL) has been setup to enable component testing using HIL approaches as well as holistic and system-level testing \cite{SGTL}. In the past, Power-Hardware-in-the-loop component testing (PHIL) of Photovoltaic inverters and distributed series reactors took place in the lab \cite{Diss_Alfio}, \cite{Pohl}.

In order to efficiently control low-voltage grids, observability and automation are necessary prerequisites. Conventionally, low-voltage grids have been treated as power sinks without any significant control possibilities. While there were extremely low to no measurements in the distribution grids of the past, the evolution of smart meters and sensors have made the grids more observable. Due to the number of low-voltage grids as well as the number of DERs and new loads such as electric vehicles and heat pumps, the amount of controllable components has increased manifold and led the way towards smart grids. Historically, while a number of contributions in the literature deal with developing automation systems for smart grid applications, only a few of them were dedicated to experimentally testing the algorithms. Testing smart grid applications with laboratory experiments and field tests has been an important research field lately. There have been several projects in the distribution grid dealing with the topic of congestion management \cite{Exner} - \cite{Lysikatos}.

One of the drawbacks of most of the proposed solutions in the literature is that almost all of the control applications are proprietary solutions and not modular. That means that the software is specific to a particular hardware, and updating or integrating new algorithms from other users becomes an arduous task every time. To enhance the feasibility of interoperability, previous work at the TU Dortmund University dealt with creating algorithms based on software using a standard developed for the automation of distribution grid substations \cite{World}, \cite{ISGT}. Thus, the control algorithms are configured using data models according to IEC 61850-7-3 \cite{61850-7-3} and IEC 61850-7-4 \cite{61850-7-4}. This offers modularity and the possibility of multiple users creating various applications that can essentially work together on the same hardware device. The novel system architecture can aggregate various protection and control functions onto industrial hardware platforms, thereby accentuating the need for hardware independence, which makes this research the first of its kind. This paper deals with the implementation of one such control algorithm and experimentally verifies it in a low-voltage test grid. 
Taking the previous introduction on system-level testing and hardware-independent smart grid applications into account, the research questions of this paper are as follows: 
\begin{itemize}
    \item[A)] How can hardware-independent control algorithms be designed and validated in close-to real-world conditions?
    \item[B)] Can congestions be effectively managed curatively using an Optimal Power Flow (OPF) algorithm in low-voltage grids?
    \item[C)] Can redispatch be curatively provided from low-voltage grids to the higher grid level without the knowledge of the future grid state? 
\end{itemize}
The aims of this paper are to test and evaluate the capability of a conventional OPF algorithm for congestion management in an exemplary low-voltage grid on a system level. In \cite{1}, applications on the system level are described as applications which influence the operation of the power system, which fits the test presented in this paper. The communication and control of several DERs using modern control approaches are often data-driven control approaches or require predictions of the future grid state \cite{Data_Driven}, \cite{MPC}. Due to the lack of data and forecasts of future grid states in low-voltage grids, this paper evaluates whether classical control approaches are suitable for usage in low-voltage grids. Consequently, the contribution of this paper is as follows: 
\begin{itemize}
    \item The hardware-independent smart grid control application (SG App) is presented and is tested with a low-voltage test setup in the SGTL at TU Dortmund University,
    \item A modified OPF implemented on the SG App is presented and used together with a Weighted-Least-Square-Value SE to monitor and control the low-voltage grid,
    \item The capability of the OPF algorithm to perform congestion management and curatively provide redispatch at the medium-voltage-low-voltage transformer is evaluated.  
\end{itemize}
\begin{figure}[t]
\centerline{\includegraphics{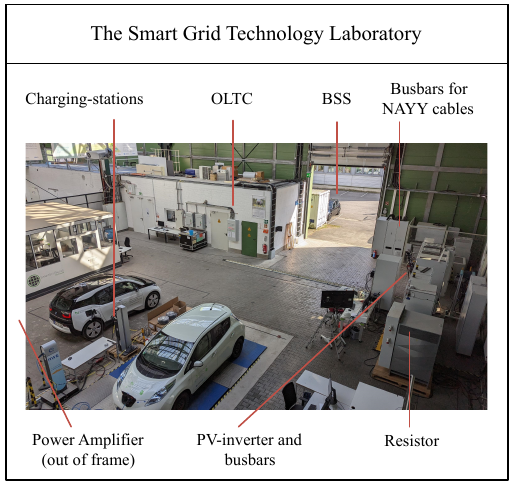}}
\caption{The components of the SGTL}
\label{fig_Lab}
\end{figure}

The remainder of this paper is as follows: Section II describes the current state of the proposed hardware-independent software research at TU Dortmund University. Section III explains the test setup and test case of the given experiment. Section IV explains the modified OPF implementation. While Section V explains the results, the paper ends in Section VI with concluding remarks.

\section{The hardware-independent Smart Grid Application}
\begin{figure}[t]
\centerline{\includegraphics{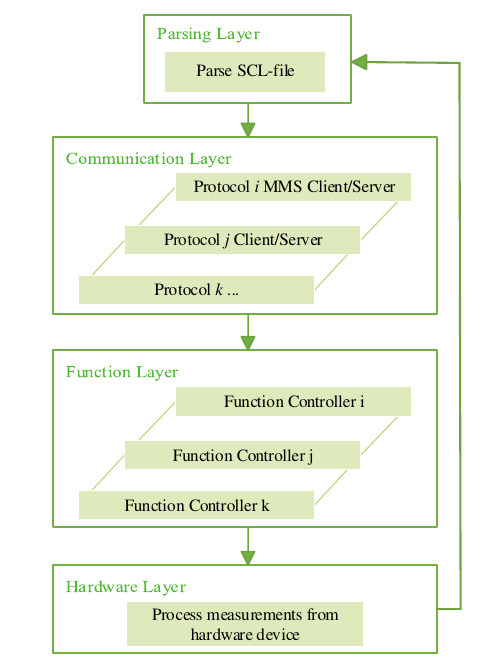}}
\caption{Structure of the smart grid application}
\label{fig_SG_App}
\end{figure}

As already mentioned, the biggest problem among the proposed solutions for distribution grid automation in the literature is that most of the applications are bound to some kind of hardware. In order to avoid this problem, a system architecture enabling distribution grid automation in the form of higher-level and coordinated control functions for grid monitoring and control, executed on various industrial hardware devices such as \cite{Kocos} or more generic embedded computers such as raspberry pi, has been developed \cite{Björn}. To achieve this, the SG App has been developed in several layers: The Parsing Layer, the Communication Layer, the Function Layer and the Hardware Layer. The structure of the SG App is depicted in Fig. \ref{fig_SG_App}. In the Parsing Layer, a configuration file according to the Substation Configuration Language (SCL) specified in IEC 61850-6, including the intelligent electronic devices (IEDs), substations and the communication is parsed into an online data model \cite{61850-6}. In the Communication Layer, the communication between the different IEDs according to various protocols is initialised, and the data is communicated. Communication protocols include Transfer Control Protocol (TCP), Modbus-TCP, the IEC 60870-5 protocol and IEC 61850-8 based MMS \cite{61850-8-1}, \cite{104}. The Function Layer contains the function controllers, the function implementation itself and input/output layers for every function assigned to a specific IED running the application. Thus, distributed control architectures, including several devices running the SG App with the same SCL file, are possible. Finally, the Hardware Layer specific to different hardware devices is called, where measurements from the hardware device are processed and written to the respective logical nodes specified in the SCL file.

\section{The test case and test specification}
In this section, the test case, the control algorithm under test and the test specification are described based on the Holistic Test Description given in \cite{1}. The applied procedure is a proposed guideline to standardise test descriptions and documentation in power system testing and is adopted in this publication. A guideline for the usage of the proposed test description is given in \cite{2}, \cite{De Paola}.
\subsection{The test case}\label{subsec_test_case}
\begin{figure}[t]
\centerline{\includegraphics{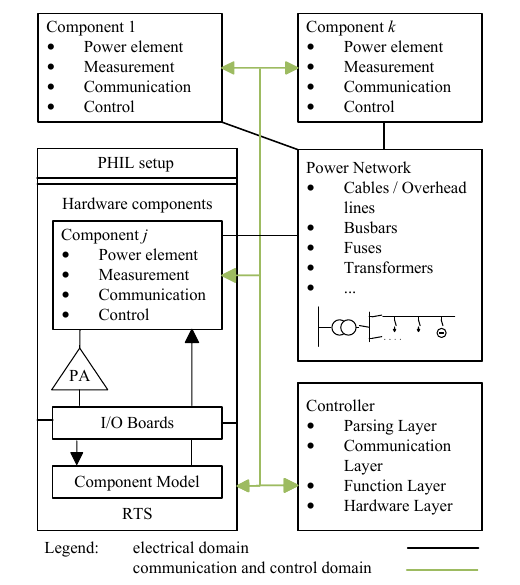}}
\caption{Structure of the given laboratory test setup}
\label{fig_lab_testing}
\end{figure}

The test objective is to test the capability of an OPF to cure congestions and to provide curative redispatch to the DSO in the form of adapting the apparent power flow over the transformer at any given time if it exceeds predefined limits. Thus, the objective function of the OPF is altered, and time-variant constraints are included. The SE is used to monitor the grid state, and the OPF is used to control the operational behaviour of its components, considering the goals specified above. The object under investigation is thus the OPF algorithm developed in the previously described software. The OPF is used to control several hardware devices, including a PHIL-controlled PV inverter, in a configurable low-voltage test grid. Thus, the OPF algorithm and the behaviour of the hardware devices are jointly tested. The test setup of this paper is depicted in Fig. \ref{fig_lab_testing}.

The system under test consists of the components and subsystems depicted as part of the research infrastructure facility Fig. \ref{fig_Lab}, whilst the object under test is the controller. The components of the system are described in the following and in \cite{SGTL}, \cite{Alfio_PHIL}: 
\begin{itemize}
\item $10/0.4$ kV On-Load-Tap-Changer (OLTC) 
\item Redox-Flow Battery Storage System (BSS) with a nominal power $S_{\mathrm{max}}^{(3)} = 30$ kVA and capacity $E_{\mathrm{total}} = 100$ kWh
\item PHIL system consisting of a PV inverter with a nominal power $P_{\mathrm{max}}^{(3)} = 60$ kVA, a RTS and a Power-Amplifier with a nominal output power of $200$ kVA 
\item Controllable Resistor with a maximal load of $P_{\mathrm{max}} = 200$ kW
\item Charging-station (CS) and EVs with charging current limits $I_{\mathrm{max}} = 16$ A and $I_{\mathrm{min}} = 6$ A. 
\item Power-quality-measurement-devices at busbar 007 and busbar 008
\item NAYY-J low-voltage cables with a cross-section of $25~\mathrm{mm^2}$ and $\mathrm{150~mm^2}$ 
\end{itemize}

\begin{figure*}[t]
\centerline{\includegraphics[scale=1.1]{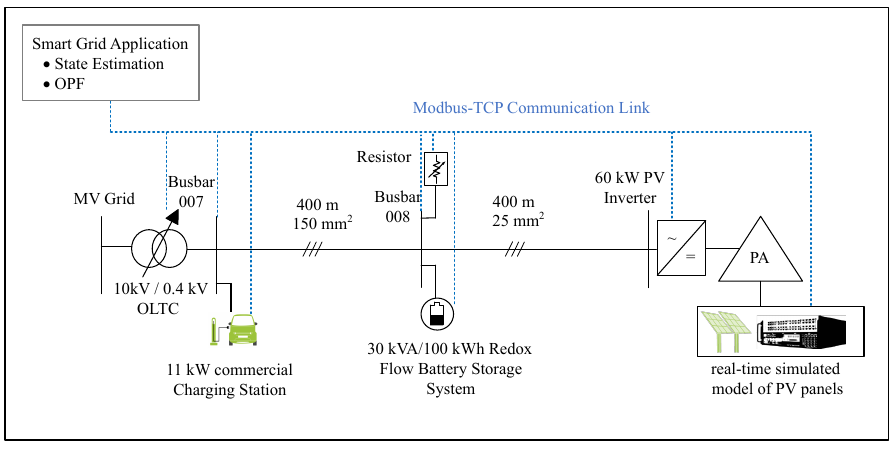}}
\caption{The laboratory set up in the SGTL}
\label{fig_Topology}
\end{figure*}

\begin{table}[t]
\caption{Data obtained by the assets in the SGTL}
\begin{center}
\begin{tabular}{|c|c|c|}
\hline
\textbf{Asset} & \textbf{Communicated Data} & \textbf{R/W)} \\
\hline
OLTC & tap position $\tau$  & R/W \\
     & phase to ground voltage $V$ & R  \\
     & tap position limits $\tau_{min}$, $\tau_{max}$  & R\\ 
     & voltage change per step & R \\
\hline
PV inverter & phase to ground voltage $V$ & R \\
			& three phase power $P^{(3)}$, $Q^{(3)}$ & R \\
			& three phase power setpoint $P^{(3)}_{\mathrm{set}}$, $Q^{(3)}_{\mathrm{set}}$ & R/W \\
			& maximum three phase power $P^{(3)}_{\mathrm{max}}$ & R \\
\hline
CS                  & charging current $I$ & R/W \\
					& charging power $P$ & R \\
					& current limits $I_{\mathrm{min}}$, $I_{\mathrm{max}}$ & R \\
					& vehicle state according to IEC 61851 \cite{IEC 61851}& R \\
					\hline 
BSS & State of Charge (SoC) $E_{t_0}$ & R \\
	& SoC limits $E_{\mathrm{max}}$ and $E_{\mathrm{min}}$ & R \\
	& phase to ground voltage $V$ & R \\
	& current $I$ & R \\
	& active, reactive and apparent power $P$, $Q$, $S$ & R \\
	& three phase setpoints $P^{(3)}_{\mathrm{set}}$, $\mathrm{cos}(\phi)^{(3)}_{\mathrm{set}}$ & R/W \\
	& maximum power $S^{(3)}_{\mathrm{max}}$ & R \\
\hline 
measurement  & current $I$ & R/W \\ 
devices		     	& phase to ground voltage $V$ & R \\
				& active, reactive and apparent power $P$, $Q$, $S$ & R \\	
				& power factor $\mathrm{cos}(\phi)$ & R \\
\hline
RTS & Temperature $T$ in $^\circ C$ & R\\
    & Irradiation $E$ in $\frac{W}{m^2}$ & R \\
    & Synchronisation bit & R \\ \hline 
\end{tabular}
\label{table:data}
\end{center}
\end{table}

The components are electrically connected in the testbed using busbars in the cabinets. There it is possible to flexibly arrange the assets using underground cables and additional busbars. As observable in Fig. \ref{fig_lab_testing}, the domains under investigation are not only the electric domain but also communication and control. The communication link is based on Modbus-TCP, whereas the components act as Modbus servers while the SG App acts as Modbus client \cite{Modbus}. The data that is being communicated is given in Table \ref{table:data}. Three phase daata of the electrical grid is considered to be symmetrical. In order to perform an OPF, nodal powers for each node in the grid are required. The SE is commonly used in the transmission grid to identify the exact grid state based on field measurements with a high amount of measurement redundancy. In distribution grids, several approaches of estimating the grid state are finding more and more usage to obtain the grid state in previously unobservable grids. As using a state estimation algorithm to estimate the grid with measurement sparsity is not the focus of this paper, it is assumed that the grid is observable or pseudo-observable, i.e. $\eta = \frac{2n-1}{m} \geq 1$ holds, where $n$ is the number of nodes and $m$ the number of measurements. The SE is thoroughly described in the literature and will not be further described in this paper, except that the SE used in this approach is the classic Weighted-Least Squares approach \cite{SE}.

The OPF is a mathematical optimization technique which is commonly used in power system analysis and operation. The general formulation is given in equations (\ref{eq_OPF_1}) to (\ref{eq_OPF_n}) \cite{Matpower}. Use cases of the OPF in power systems are the economic dispatch of power plants on the transmission grid level or the provision of ancillary services such as voltage and frequency control \cite{OPF_dispatch}, \cite{OPF_voltage}. In this paper, the OPF is used for curative congestion management and the provision of redispatch. The concrete specification of the OPF follows in section \ref{sec_OPF}.

\begin{equation}
\min_x f(x) \label{eq_OPF_1}
\end{equation} 
subject to 
\begin{align}
\label{eq_OPF_2} &g(x) = 0    \\
\label{eq_OPF_3} &h(x) \leq 0   \\
\label{eq_OPF_n} &x_{\mathrm{min}} \leq x \leq x_{\mathrm{max}}
\end{align}

The repeatability of the tests is ensured by synchronising the first execution of the SE and the start of the resistor's load profile given in Fig. \ref{fig_Widerstand} with the start of the model of the RTS. This is realised by deploying a virtual Modbus server and waiting for the RTS to write in a specific Modbus register. The test criteria are the capability of the OPF to cure voltage violations, apparent power-flow violations over lines, and to keep the apparent power-flow over the transformer within the predefined limits. Moreover, the successful execution of a test using the SG App in an embedded system is validated. The target metrices are given in equations (\ref{eq_voltage_violations}), (\ref{eq_power_violations}), (\ref{eq_amount_violated_U}) and (\ref{eq_amount_violated_S}). The total number of times that voltage violations $N_v$ and thermal violations $N_s$ could not be cured is an indicator of the capability of the OPF to cure violations in the grid. The variable $n_v(t)$ is $1$ for every time step $t$ in which a previous violation could not be cured and is defined in equation (\ref{eq_nv_ns}), the definition of $n_s(t)$ is done accordingly. 
\begin{align}
\label{eq_voltage_violations} N_v = \sum_{t=2}^{T} n_v(t)\\
\label{eq_power_violations}N_s = \sum_{t=2}^{T} n_s(t) 
\end{align}
\begin{equation}
\label{eq_nv_ns} n_v(t) = \begin{cases}
1,~\text{voltage violation occurs in } t-1\text{ and } t \\
0,~\text{otherwise}\end{cases}
\end{equation}
The second test metric is given in equations (\ref{eq_amount_violated_U}) and (\ref{eq_amount_violated_S}). The sum of violations over all respective violations in the timesteps $t~\in ~\{t_0,...,T\}$ and nodes $i~\in ~\{1,...,n\}$ after a congestion took place in the previous timestep is calculated. The superscript "control" depicts the controlled grid state, while "reference" depicts the uncontrolled reference grid state. In order for the OPF to successfully cure voltage violations, power-flow violations and to provide redispatch to the upper grid level, this value has to be close to $0$. A qualitative assessment of the results of the OPF follows in section \ref{sec_5}. 
\begin{align}
\label{eq_amount_violated_U} A_v = \sum_{t=2}^T \sum_{i=1}^n n_v(t) \cdot |V_i(t)^{\mathrm{control}}-V_i(t)^{\mathrm{reference}}|  \\
\label{eq_amount_violated_S} A_s = \sum_{t=2}^T \sum_{i=1}^n n_s(t) \cdot |S_{ij}(t)^{\mathrm{control}}-S_{ij}(t)^{\mathrm{reference}}| 
\end{align} 
\subsection{The test specification}
\begin{figure}[t]
\centerline{\includegraphics{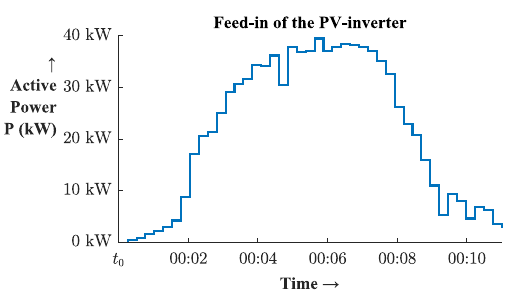}}
\caption{The time series of the PV inverter}
\label{fig_PV_Zeitreihe}
\end{figure}

\begin{figure}[h]
\centerline{\includegraphics{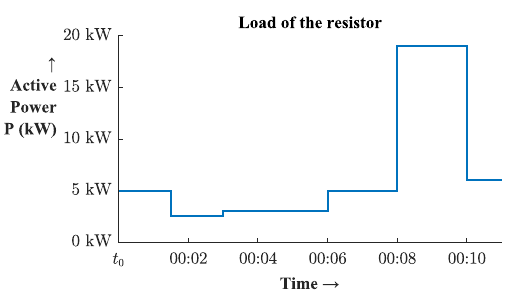}}
\caption{The time series of the resistive load}
\label{fig_Widerstand}
\end{figure}

The system under test is presented in Fig. \ref{fig_Topology}. There, two $400$ m long cables are used to depict a long feeder as it typically occurs in rural grids \cite{Simbench}. At the end of the feeder, the PV inverter feeds in active power corresponding to irradiation data of a sunny day as presented in Fig. \ref{fig_PV_Zeitreihe}. The test takes $11$ minutes as the PV-model on the RTS presented in \cite{Diss_Alfio} reassembles the feed-in of a day in $11$ minutes. The test is synchronized with the start of the PHIL control of the PV inverter to achieve reproducible testing. The SE and OPF are executed every 15 seconds, with $t_0 =$ 00:00 and $T =$ 00:11. At busbar 008, the resistor imposes a load on the grid as presented in Fig. \ref{fig_Widerstand}. The battery is connected to busbar 008 through a NYY-J 5x16 RE cable with approximated length of 100 m and is operating at a SoC of $50$ \% and a feed-in of $0$ kW. The charging station is connected to busbar 007 with a cable of type H07RN-F 5G6 and a length of 35 m, a car will be plugged in directly after $t =$ 00:05:45.
The measurement data that is obtained is given in Table \ref{table:data}. Measurement equipment installed in the busbars is used as well as the internal measurements from the BSS, the CS, the PV inverter and the OLTC. These measurements make the grid observable. The output of the SE is the system state consisting of nodal voltages, and power-flows as given in (\ref{eq_system_state}).
\begin{equation}
x = [V_i,\delta_i,P_i,Q_i]\quad \forall ~ i \in \lbrace 1,...,n \rbrace 
\label{eq_system_state}
\end{equation}
The system state (\ref{eq_system_state}) serves as an input to the OPF. The output of the OPF and cascaded control of the OLTC are setpoints $P_{\mathrm{set}}$ and $Q_{\mathrm{set}}$ for active and reactive power of the PV inverter and the BSS, for the active power of the EV, as well as the tap position $\tau_{\mathrm{set}}$ of the OLTC. The control of the OLTC is executed in the controller of the OPF function and is not part of the OPF itself. The current grid state is observed and if only a voltage violation is detected and the OLTC has not been stepped in the last time instance of the controller, a tap change $\tau_{\mathrm{t}} = \tau_{\mathrm{t-1}} \pm 1$ will be executed instead of the optimization. For the EV, only the active power is controllable by setting the charging current $I^{\mathrm{set}}$ as an integer between $6$ A and $16$ A. 

To specify the occurrence of voltage violations, power-flow violations and the provision of redispatch in the given test, limits $S_{\mathrm{max}}$, $S_{\mathrm{min}}$, $V_{\mathrm{max}}$ and $V_{\mathrm{min}}$ have to be specified for every line or respectively node. The voltage limits $V_{\mathrm{max}}$ and $V_{\mathrm{min}}$ for every node are given in (\ref{eq_vl}) and are specified according to the norm DIN EN 50160 \cite{DIN_EN_50160}. The power-flow limits $S_{\mathrm{max }}$ and $S_{\mathrm{min}}$ for the apparent power-flows are specified in Table \ref{tab_pflimit}. It is observable that the apparent power-flow over the line (Busbar 007, Busbar 008) has been artificially reduced to $40$ kVA from initially $170.4$ kVA. Moreover, due to congestion in the higher grid levels, between the time steps $t_1 =$ 00:03:30 and $t_2 =$ 00:05:00, the feed-in of the low-voltage grid to the medium-voltage grid has been restricted to $15$ kVA. 

\begin{equation}
\label{eq_vl}
0.9~\mathrm{p.u.} \leq V_i \leq 1.1~\mathrm{p.u.}\quad \forall ~ i ~ \in \{1,...,n\}
\end{equation}

\begin{table}[t]
\centering
\caption{Absolute value of apparent power-flow limits for the lines and OLTC in kVA}
\begin{tabular}{|c|c|c|}
\hline
 $t$ & (Busbar 007, MV Grid) & (Busbar 008, Busbar 007)   \\ \hline
[00:00, 00:03:30]     & $70$ &$40$ \\ \hline
[00:03:45, 00:05]    & $15$ & $40$  \\ \hline
[00:05:15, 00:11]   &  $70$  &  $40$   \\ \hline       
\end{tabular}
\label{tab_pflimit}
\end{table}

\section{The modified Optimal Power Flow Algorithm}\label{sec_OPF}
The optimization vector
\begin{equation*}
x = [V_i,\delta_i,P_i,Q_i]\quad \forall ~ i  ~ \in \lbrace 1,...,n \rbrace
\end{equation*}
consists of complex nodal voltages and powers for every node. The objective function is given in (\ref{eq_objective}). It aims to minimise the difference of the active power setpoint to a target value of the active power for every flexibility as well as minimising the reactive power feed-in of all the assets. The variable $N_f$ denotes the number of flexibilities.
\begin{equation}
\min f(P,Q) = \sum_{i=1}^{N_f} [c_i^{\mathrm{P}}\cdot (P_i - P_i^{\mathrm{target}})^2+c_i^{\mathrm{Q}}\cdot Q_i^2]\label{eq_objective}
\end{equation}
The cost factors $c_i^{\mathrm{P}}$ and $c_i^{\mathrm{Q}}$ for different assets are described in Table \ref{tab_costs}. The costs are purely of qualitative nature and do not depict the real costs of curtailing these flexibilities. The weighing of cost factors follows the following principles: 
\begin{itemize}
\item Reactive power should be cheaper than active power. 
\item The battery storage system should be the cheapest flexibility. 
\item The curtailment of the PV inverter is cheaper than the curtailment of the charging process of the EV, as it does not influence the customer comfort. 
\end{itemize}
\begin{table}[t]
\caption{Cost factors for the different flexibilities as well as target values for active power}
\begin{center}
\begin{tabular}{|c|c|c|c|}
\hline
\textbf{Flexibility} & \textbf{$c_i^{\mathrm{P}}$} & \textbf{$c_i^{\mathrm{Q}}$} & $P^{\mathrm{target}}$\\
\hline
BSS & 100 & 1 & $\left\{
\begin{array}{ll}
|\frac{E_{t_0}-E_{\mathrm{total}}}{\Delta T}|,  & E < 50~\% \cdot E_{\mathrm{total}} \\
-|\frac{E_{t_0}-E_{\mathrm{total}}}{\Delta T}|, & E \geq 50~\% \cdot E_{\mathrm{total}}
\end{array}
\right. $\\ \hline
PV & 1000 & 10 & $-P_{\mathrm{max}}^{(3)}$ \\ \hline
EV & 10000 & $\infty$ & $P_{\mathrm{max}}=I_{\mathrm{max}}\cdot V_i \cdot \sqrt{3}$\\ \hline
\end{tabular}
\label{tab_costs}
\end{center}
\end{table}
In the following, the target values for active power of the flexibilities are explained. The aim is to minimise the influence on the operational behaviour of the flexibilities. Thus, if an EV is connected, $P^{\mathrm{target}}$ is the maximum charging power for the EV, otherwise $P^{\mathrm{target}}_{\mathrm{EV}} = 0$ W holds. The information, if an EV is connected and ready for charging, is obtained from the CS. The calculation of $P^{\mathrm{target}}_{\mathrm{PV}}$ is more complex. The currently available power for the PV inverter is a very important quantity, as without it, a proper upper bound for the respective inequality constraint cannot be formulated. This can either be obtained by the inverter providing this data or by having irradiance and temperature measurements. The given inverter in the SGTL does not communicate the maximally available active power. Thus, the irradiance and temperature of the emulated PV module of the PHIL setup presented in Fig. \ref{fig_Topology} are used to calculate the maximum available power of the inverter according to equation (\ref{eq_PV_Inverter_Power}). The variable $T$ denotes the temperature in $^{\circ}C$, the variable $E$ denotes the irradiation in $\frac{\mathrm{W}}{\mathrm{m}^2}$ and the variable $\alpha$ the temperature coefficient of the respective PV module in $\frac{\mathrm{W}}{\mathrm{K}}$. 
\begin{equation}
P_{\mathrm{max}}^{\mathrm{PV}}(E,T,\alpha) =  P_{\mathrm{ref}} \cdot \frac{E}{E_{\mathrm{ref}}} \cdot (1+\alpha \cdot (T-T_{\mathrm{ref}})) \cdot \eta_{\mathrm{inverter}}\label{eq_PV_Inverter_Power} 
\end{equation}
The temperature coefficient $\alpha$ has been calculated in a test case with $T = 45~ ^{\circ}C$, $E = 1000~\frac{\mathrm{W}}{\mathrm{m^2}}$ and is $\alpha =  0,00273~\frac{W}{K}$. The efficiency $\eta$ of the inverter has been calculcated using the European efficiency as given in (\ref{european_efficiency})\cite{eq_european_efficiency}.
\begin{align}
\eta_{\mathrm{inverter}} &= 0.03 \cdot \eta_{\mathrm{5 \%}} + 0.06 \cdot \eta_{\mathrm{10 \%}} + 0.13 \cdot \eta_{\mathrm{20 \%}} + 0.10 \cdot \eta_{\mathrm{30 \%}}  \nonumber\\
& + 0.48 \cdot \eta_{\mathrm{50 \%}} + 0.20 \cdot \eta_{\mathrm{100 \%}}
\label{european_efficiency}
\end{align}
For the given inverter, $\eta_{\mathrm{inverter}} = 0.93$ resulted according to the measurements given in Table \ref{tab_PV_Inverter}. 
\begin{table}[t]
\begin{center}
\begin{threeparttable}
\caption{Measurements for the calculation of the efficiency of the given PV inverter}
\begin{tabular}{|c|c|c|c|c|}
\hline
\% of nominal power & $P_{\mathrm{DC}}$ in kW & $P_{\mathrm{AC}}$ in kW & $\eta_i$ \\
\hline
100 & 62.22 & 60\tnote{1} & 0.9643   \\ \hline
50 & 31.11 & 29.7 & 0.9547 \\ \hline 
30 & 18.66 & 16.6 & 0.8893 \\ \hline
20 & 12.44 & 10.8 & 0.8679 \\ \hline 
10 & 6.22 & 5.2 & 0.8357 \\ \hline 
5 & 3.11 & 2.4 & 0.7715 \\ \hline
\end{tabular}
\begin{tablenotes}
 \item[1] This measurement could not be obtained due to stability issues with the inverter and was assumed \end{tablenotes}
\label{tab_PV_Inverter}
\end{threeparttable}
\end{center}
\end{table}
The target power for the BSS is calculated using a case distinction as given in Table \ref{tab_costs}. If the SoC is below $50~ \%$, then $P^{\mathrm{target}}$ is set as $|\frac{E_{t_0}-E_{\mathrm{total}}}{\Delta T}|$, if it is above $50~ \%$, then $P^{\mathrm{target}}$ is set as $-|\frac{E_{t_0}-E_{\mathrm{total}}}{\Delta T}|$. The quantity $\frac{E_{t_0}-E_{\mathrm{total}}}{\Delta T}$ denotes the missing power for the BSS to reach a SoC of $50 \% \cdot E_{\mathrm{total}}$ in the timespan $\Delta T$. As a result, the aim of the optimization is to keep the SoC of the BSS at $50 \%$. 
The equality constraints $g$ in (\ref{eq_OPF_2}) are specified in the following equations (\ref{eq_equality_constraint_1}) to (\ref{eq_equality_constraint_2}) and describe the power-flow balance at every node $i$ in the grid. The properties $y_{ij}$ and $\theta_{ij}$ are the respective elements of the complex admittance matrix $\underline{Y}$ , $\delta_i$ and $\delta_j$ to the voltage phases of the nodes $i$ and $j$ \cite{SE}, \cite{Matpower}. The nodal powers $P_i$ and $Q_i$ are given in the consumer counting system. 
\begin{align}
\label{eq_equality_constraint_1}P_i + V_i \cdot \sum_{j = 1}^n V_j \cdot y_{ij} \cdot \mathrm{cos}(\delta_i-\delta_j-\theta_{ij}) = 0 \\
\label{eq_equality_constraint_2}Q_i + V_i \cdot \sum_{j = 1}^n V_j \cdot y_{ij} \cdot \mathrm{sin}(\delta_i-\delta_j-\theta_{ij}) = 0 
\end{align}

The inequality constraints $h$ in (\ref{eq_OPF_3}) correspond to the power-flow constraints over the lines in the grid. They are specified in equations (\ref{eq_pf1}) and (\ref{eq_pf2}). The limit of the apparent power-flow $S_{\mathrm{ij,~max}}(t)$ for a line $(i,j)$ is given as an external input and can continuously be updated by the DSO. This enables the DSO to dynamically reduce the load or feed-in from the low-voltage grid if its flexibility potential is high enough.
\begin{align}
S_{ij}(\theta_i,V_i)-S_{ij\mathrm{,~max}}(t) &\leq 0 \label{eq_pf1}\\
S_{ji}(\theta_j,V_j)-S_{ji\mathrm{,~max}}(t) &\leq 0\label{eq_pf2}
\end{align}
The second set of inequalities as given in (\ref{eq_OPF_n}) deals with voltage and power limitations and is specified in equations (\ref{eq_voltage}) to (\ref{eq_reactive_power}). The limits for active and reactive power for the BSS, CS and the PV inverter are given in Table \ref{table_Asset_limits} where a maximal $\mathrm{sin}(\phi)$ of 0.44 is assumed. 
\begin{align}
\label{eq_voltage} V_i^{\mathrm{min}} & \leq V_i \leq V_i^{\mathrm{max}}&\quad \forall ~i \in \lbrace 1,...,n \rbrace \\
\label{eq_active_power} P_{i}^{\mathrm{min}} & \leq P_{i} \leq P_{i}^{\mathrm{max}}&\quad\forall ~ i \in \lbrace 1,...,N_f \rbrace \\
-360^{\circ}& \leq \delta_i \leq 360^{\circ} &\quad\forall ~ i \in \lbrace 1,...,n \rbrace\\
\label{eq_reactive_power}Q_{i}^{\mathrm{min}} & \leq Q_{i} \leq Q_{i}^{\mathrm{max}} &\quad \forall ~ i \in \lbrace 1,...,N_f \rbrace 
\end{align}

\begin{table}[t]
\caption{Active and reactive power limits for the flexibilities}
\begin{center}
\begin{tabular}{|c|c|c|c|}
\hline
 & BSS & PV inverter & Charging Station \\ \hline
$P_{\mathrm{min}}$ & $-S_{\mathrm{max}}$ & 0 kW & $3 \cdot V_{\mathrm{CS}} \cdot I_{\mathrm{min}}$ \\\hline
$P_{\mathrm{max}}$ &  $S_{\mathrm{max}}$ & -$S_{\mathrm{max}}$ & $3 \cdot V_{\mathrm{CS}} \cdot I_{\mathrm{max}}$ \\ \hline
$Q_{\mathrm{min}}$ & $- S_{\mathrm{max}} \cdot \mathrm{sin}(\phi)$ & $- S_{\mathrm{max}} \cdot \mathrm{sin}(\phi)$ & 0 kVar \\ \hline
$Q_{\mathrm{max}}$ & $S_{\mathrm{max}} \cdot \mathrm{sin}(\phi)$ & $S_{\mathrm{max}} \cdot \mathrm{sin}(\phi)$ & 0 kVar \\ \hline
\end{tabular}
\label{table_Asset_limits}
\end{center}
\end{table}

\begin{table}[t]
\caption{Target metrices for lines and nodes}
\begin{center}
\begin{tabular}{|c|c|c|c|}
\hline
 & Uncontrolled state & Controlled state & Reduction in \%
\\ \hline
$N_v^{\mathrm{PV}}$ & 17 & 6 & 64.71 \\ \hline
$A_v^{\mathrm{PV}}$ & 0.3243 p.u. & 0.1126 p.u. & 65.28 \\ \hline
$N_s^{\mathrm{OLTC}}$ & 6 & 6 & 0 \\ \hline 
$A_s^{\mathrm{OLTC}}$ & 40.19 & 35.99 kVA & 10.45 \\ \hline
$N_s^{\mathrm{007,008}}$ & 7 & 5 & 28.57 \\ \hline
$A_s^{\mathrm{007,008}}$ & 94.30 kVA  & 26.19 kVA & 72.23 \\ \hline
\end{tabular}
\label{tab_results_N_A}
\end{center}
\end{table}

\section{Scenarios and Results}\label{sec_5}
\begin{figure}[t]
\centering
\includegraphics{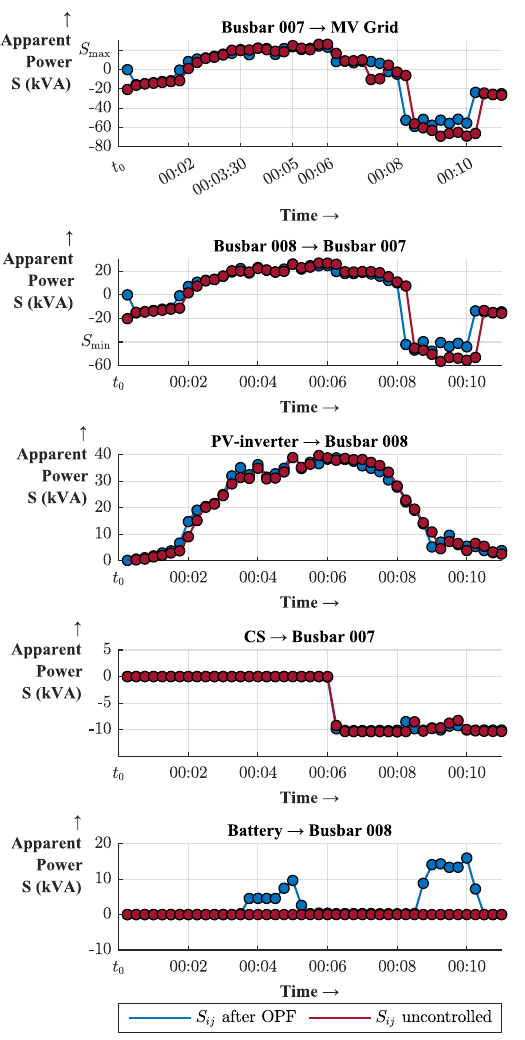}
\caption{Controlled and uncontrolled apparent power-flows in kVA}
\label{fig_OPF_S_results}
\end{figure}

\begin{figure}[t]
\centering
\includegraphics{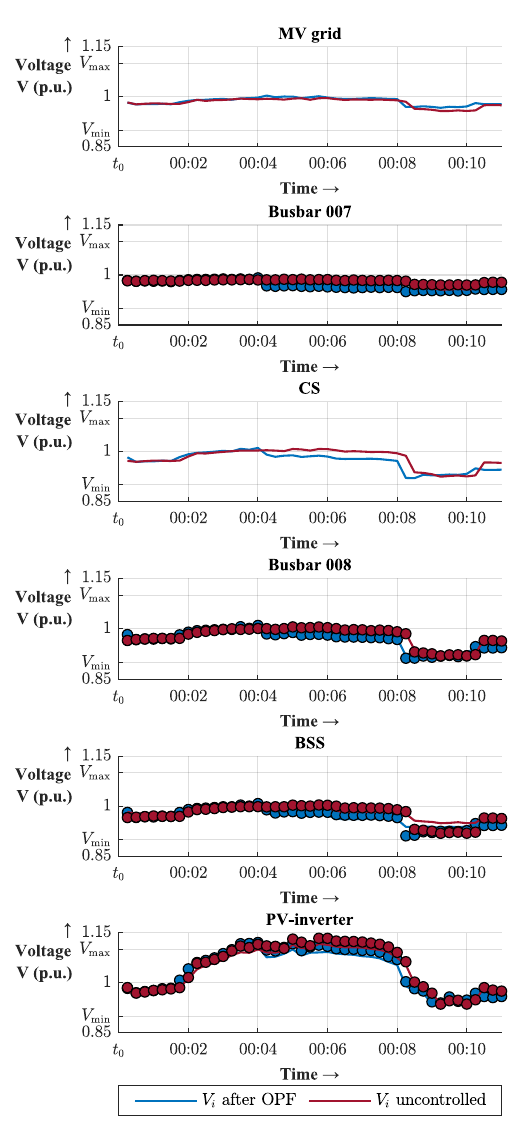}
\caption{Controlled and uncontrolled voltage in the grid in p.u.}
\label{fig_OPF_V_results}
\end{figure}

In this section, the uncontrolled system state as presented in subsection \ref{subsec_test_case} is given and compared to the controlled system state. Moreover, the target metrics (\ref{eq_voltage_violations}), (\ref{eq_power_violations}), (\ref{eq_amount_violated_U}) and (\ref{eq_amount_violated_S}) are presented and evaluated regarding the question of the capability of the OPF to cure voltage violations, power-flow violations and provide curative redispatch. The uncontrolled and the controlled grid state are presented in Fig. \ref{fig_OPF_S_results} and Fig. \ref{fig_OPF_V_results}. It is observable that without any control actions, the apparent power-flow over the line (Busbar 008, Busbar 007) exceeds  $S_{\mathrm{min}} = -40$ kVA between 00:08:30 and 00:10:15. Moreover, the temporary power-flow-limit of $15$ kVA over the transformer given between 00:03:30 and 00:05:00 is exceeded at all times in the time interval. It is observable that in the uncontrolled grid state, a voltage violation occurs between 00:03:30 and 00:07:45. 
The target metrices $N_v$,$N_s$, $A_v$ and $A_s$ for the controlled and uncontrolled grid state are presented in Table \ref{tab_results_N_A}. One can see that the OPF manages to cure the power-flow violations on the line (Busbar 008, Busbar 007) two times completely, thus $N_s^{\mathrm{007,008}}$ reducing from $7$ in the uncontrolled grid state to $5$ in the controlled grid state. 

\begin{figure}[t]
\centering
\includegraphics{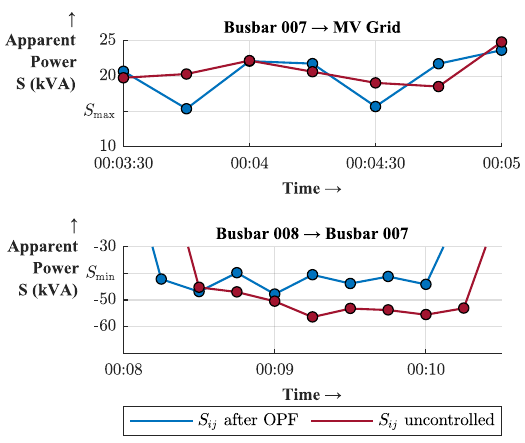}
\caption{Close-up of power.flows on (Busbar 007, MV Grid) and (Busbar 008, Busbar 007)}
\label{fig_OPF_close_results}
\end{figure}
Moreover, the amount of power-flow violations reduces by $72.23$ \%. The optimization significantly reduces the power-flow congestions over the given line, but is not able to completely cure them. A close-up in Fig. \ref{fig_OPF_close_results} in the time interval between 00:08 and 00:10 shows that an alternating pattern results: The violation is cured and $S_{\mathrm{007,008}} \approx -40$ kVA occurs. In the next time step the violation is increasing again. The load of the resistor is constant between 00:08 and 00:10 with 19 kW per phase. However, the feed-in of the PV inverter drastically reduces from approximately $30$ kVA to less than $10$ kVA in the given time interval. Thus, the loading of the grid increases and the OPF is able to reduce the power-flow violations, but not completely cure them. 

Moreover, the limit $S_{\mathrm{max}} = 15$ kVA over the OLTC in the uncontrolled grid state cannot be cured in any of the given time steps, i.e. $N_v^{\mathrm{PV}}$ reduces by $0$ \%. The power-flow violations of the given limit $S_{\mathrm{max}} = 15$ kVA can only be reduced by $10.45$\% compared to the uncontrolled grid state. Closely looking at the power-flows over (Busbar 007, MV grid) in Fig. \ref{fig_OPF_close_results} in the given interval [00:03:30,00:05:00] shows that at 00:03:45 and at 00:04:30 the violation can be cured, however, due to the increasing feed-in of the PV inverter at the time, in the next time steps $t+1 = $00:04 and 00:05, a violation occurs again. 

The third violation that can be observed in the uncontrolled grid state is the voltage violation at the PV inverter. There, in the uncontrolled grid state, $17$ voltage violations occur. The number of voltage violations is significantly reduced by $64.71$~\% to $6$ in the controlled state. Moreover, the amount of voltage violations $A_v^{\mathrm{PV}}$ reduces by $65.28$ \%. These results are obtained by a step of the transformer at 00:04:15 from position $5$ to $4$, thus reducing the voltage in the grid by $0.025$ p.u. The cascaded control of the OLTC is one of the main reasons for the effective voltage control of the OPF, as a step change of the OLTC significantly impacts the voltage in the grid. However, as it prevents controlling the different DERs of the grid, it reduces the quality of congestion management and redispatch provision. 

\section{Conclusion and Outlook}
While various papers in the literature have dealt with the provision of congestion management in the distribution grids, this paper evaluated the same along with the provision of curative redispatch capability of an OPF algorithm in a low-voltage test grid. The test setup enables the testing of grid monitoring and control algorithms in a real low-voltage grid in a controlled environment. The OPF is able to reduce power flow and voltage congestions by $72.23$ \% and $64.71$ \% in the given test case. Thus, complete curing of congestions is not possible and the capability of the OPF to cure congestions can only be partially verified. Research question B) dealt with the question of curative redispatch without taking the future grid state into consideration. Here, it is observable that the OPF is not able to guarantee given apparent power flows over the transformer, the violation of the predefined power-flow trajectory can only be reduced by $10.45$ \% compared to the uncontrolled grid state. At last, the successful validation of a hardware-independent control software using the SE and OPF algorithms in a low-voltage test grid was shown. In future work, preventive control algorithms such as Model Predictive Control will be used and their capability to provide congestion management and redispatch will be evaluated in the presented test grid, whilst considering varying quality of forecast data for the future grid state.


\begin{thebibliography}{00}
\bibitem{Breker} S. Breker, A. Claudi and B. Sick, "Capacity of Low-Voltage Grids for Distributed Generation: Classification by Means of Stochastic Simulations", IEEE Transactions on Power Systems, vol. 30, no. 2, pp. 689-700, 2015
\bibitem{Hu} J. Hu, C .Ye and Y. Ding, "A Distributed MPC to exploit reactive power V2G for real-time voltage regulation in distribution networks", IEEE Transactions on Smart Grids, vol. 13, no. 1, pp. 576-588, 2022
\bibitem{Montoya} J. Montoya et al., “Advanced Laboratory Testing Methods Using Real-Time Simulation and Hardware-in-the-Loop Techniques: A Survey of Smart Grid International Research Facility Network Activities”, Energies, vol. 13, no. 12, p. 3267, Jun. 2020.
\bibitem{Strasser} T. Strasser, M. Stifter, F. Andren et al., "Co-Simulation Training Platform for Smart Grids", IEEE Transactions on Power Systems, vol. 29, no..4 , pp. 1989-1997, July 2014
\bibitem{1} K. Heussen, D. Babazadeh, M. Z. Degefa et al., Test Procedure and Description for System Testing, "European Guide to Power System Testing: The ERIGrid Holistic Approach for Evaluating Complex Smart Grid Configurations", pp. 13-33, 2020.
\bibitem{SGTL} A. Spina et al., "Smart Grid Technology Lab – A Full-Scale Low Voltage Research Facility at TU Dortmund University", in AEIT International Annual Conference Bari, Italy, 2018.
\bibitem{Diss_Alfio} A. Spina, "Advanced laboratory testing of smart grid applications with power hardware-in-the-loop approach", Ph.D. dissertation, $\mathrm{ie}^3$, TU Dortmund University, Dortmund, 2021.
\bibitem{Pohl} O. Pohl, A. Spina, U. Häger, "Demonstration of automated curative power flow control with a multi-agent system using distributed series reactors in a PHIL simulation", Electric Power Systems Research, Volume 211, 2022
\bibitem{Brandl} R. Brandl, J. Montoya, D. Strauss-Mincu and M. Calin, "Power System-in-the-Loop testing concept for holistic system investigations", 2018 IEEE International Conference on Industrial Electronics for Sustainable Energy Systems (IESES), Hamilton, New Zealand, 2018, pp. 560-565.
\bibitem{Villarejo} M. Barragán-Villarejo, F. de P. García-López, A. Marano-Marcolini, and J. M. Maza-Ortega, “Power System Hardware in the Loop (PSHIL): A Holistic Testing Approach for Smart Grid Technologies”, Energies, vol. 13, no. 15, p. 3858, Jul. 2020.
\bibitem{Exner} C. Exner, M-A. Frankenbach, A. von Haken, N. Hübner, D.M. Gonzalez and J.P Kemper, “Findings on the practical implementation of congestion management in low- and medium-voltage networks in the flexQgrid project”, ETG Congress, Kassel, Germany, 2023.
\bibitem{Selinger-Lutz} O. Selinger-Lutz, I. Katz, R. Hollinger and C. Wittwer, "Lessons Learned From Field Test Data of a Robust Smart Grid Concept", 2018 IEEE PES Innovative Smart Grid Technologies Conference Europe (ISGT-Europe), Sarajevo, Bosnia and Herzegovina, 2018, pp. 1-6.
\bibitem{Soares} A. Soares et al., "Distributed Optimization Algorithm for Residential Flexibility Activation—Results From a Field Test", in IEEE Transactions on Power Systems, vol. 34, no. 5, pp. 4119-4127, Sept. 2019.
\bibitem{Knezovic} K. Knezović, S. Martinenas, P. B. Andersen, A. Zecchino and M. Marinelli, "Enhancing the Role of Electric Vehicles in the Power Grid: Field Validation of Multiple Ancillary Services", in IEEE Transactions on Transportation Electrification, vol. 3, no. 1, pp. 201-209, March 2017.
\bibitem{Lysikatos} I. Kouveliotis-Lysikatos, D. I. Koukoula, A. L. Dimeas and N. D. Hatziargyriou, "Plug-and-Play Algorithms for the Efficient Coordination of Active Distribution Grids", in Proceedings of the IEEE, vol. 110, no. 12, pp. 1927-1939, Dec. 2022.
\bibitem{World} R. Palaniappan et al., “Experimental verification of smart grid control functions on international grids using a real‐time simulator”, IET Generation, Transmission and Distribution, vol. 16, no. 13, pp. 2747–2760, May 2022.
\bibitem{ISGT} R. Palaniappan, B. Bauernschmitt, D. Hilbrich and C. Rehtanz, "An Intelligent Measurement and Control Device for Active Distribution Grids", 2020 IEEE PES Innovative Smart Grid Technologies Europe (ISGT-Europe), The Hague, Netherlands, 2020, pp. 975-979.
\bibitem{61850-7-3} IEC 61850-7-3:2010, AMD1:2020, International Electrotechnical Commission, 2020.
\bibitem{61850-7-4} IEC 61850-7-4:2010, AMD1:2020, International Electrotechnical Commission, 2020.
\bibitem{Data_Driven} D. Cao, J. Zhao, W. Hu et al., "Deep Reinforcement Learning Enabled Physical-Model-Free Two-Timescale Voltage Control for Active Distribution Systems", IEEE Transactions on Smart Gris, vol 13. no. 1, January 2022
\bibitem{MPC} S. Maharjan, A. M. Khambadkone, J. Chih-Hsien Peng, "Robust Constrained Model Predictive Voltage Control in Active Distribution Networks", IEEE Transactions on Sustainable Energy, vol. 12, no. 1, January 2021
\bibitem{Kocos} KoCoS Messtechnik AG: EPPECX: Power quality measurement
device. https://www.kocos.com/electrical-metrology/power-qualityanalysers/
eppe-power-quality-analysers/eppe-cx (accessed June 15, 2023). 
\bibitem{Björn} B. Bauernschmitt, R. Palaniappan, D. Hilbrich and C. Rehtanz, "Modular Configurable and Testable Automation Architecture for future Active Electrical Energy Grids", 2018 53rd International Universities Power Engineering Conference, Glasgow, UK, 2018, pp. 1-6.
\bibitem{61850-6} IEC 61850-6:2010, AMD1:2020, International Electrotechnical Commission, 2020.
\bibitem{61850-8-1} IEC 61850-8-1:2010, AMD1:2020, International Electrotechnical Commission, 2020.
\bibitem{104} IEC 60870-5-104:2006, International Electrotechnical Commission, 2006
\bibitem{2} P. Raussi et al., "Energy Systems Test Case Discovery Enabled by Test Case Profile and Repository", 2023 Open Source Modelling and Simulation of Energy Systems (OSMSES), Aachen, Germany, 2023, pp. 1-6.
\bibitem{De Paola}A. De Paola et al., "Testing-Oriented Development and Open-Source Documentation of Interoperable Benchmark Models for Energy Systems", in IEEE Open Journal of the Industrial Electronics Society, vol. 4, pp. 42-51, 2023
\bibitem{Alfio_PHIL} A. Spina, C. Rehtanz, "Power hardware-in-the-loop testbeds for advances laboratory testing of smart grid applications", Automatisierungstechnik, vol. 70, no. 12, 2022, pp. 1034-1046.
\bibitem{Modbus} Modbus Organization, "Modbus Application Protocol Specification", Version 1.1b3, 2012, \url{https://modbus.org/docs/Modbus_Application_Protocol_V1_1b3.pdf}, (accessed June 15, 2023).
\bibitem{IEC 61851} IEC 61851-1:2017, Electric vehicle conductive charging system - Part 1: General requirements, International Electrotechnical Commission, 2017
\bibitem{SE} A. Abur, A. Exposito, "Power System State Estimation: Theory and Implementation", CRC Press, 2004.
\bibitem{Matpower} R. D. Zimmerman, C. E. Murillo-Sanchez, and R. J. Thomas, “MATPOWER: Steady-State Operations, Planning and Analysis Tools for Power Systems Research and Education”, IEEE Transactions on Power Systems, vol. 26, no. 1, pp. 12–19, Feb. 2011.
\bibitem{OPF_dispatch} B.H. Chowdhury, S. Rahman, "A Review of recent advances in economic dispatch", IEEE Transactions on Power Systems, Vol. 5, No. 4, Nov. 1990.
\bibitem{OPF_voltage} V. R. Vinnakota, B. Brewer, D. Atanackovic et al., "Implementation of OPF as supportive System Voltage Control Tool in Real-time for British Columbia Transmission Network", IEEE PES General Meeting, July 2010.
\bibitem{Simbench} S. Meinecke et al., “SimBench—A Benchmark Dataset of Electric Power Systems to Compare Innovative Solutions Based on Power Flow Analysis”, Energies, vol. 13, no. 12, p. 3290, Jun. 2020.
\bibitem{DIN_EN_50160} Voltage characteristics of electricity supplied by public distribution networks, EN 50160:2010, CENELAC European Committee for Electrotechnical Standardization, 2010.
\bibitem{eq_european_efficiency} M. Fedkin, J.A. Dutton, "Efficiency of Inverters", e-Education Institute, Penn State University, \url{https://www.e-education.psu.edu/eme812/node/738}, (accessed July 15, 2023).

\end{thebibliography}
\end{document}